\documentclass[prd,aps,nofootinbib,superscriptaddress]{revtex4}
\usepackage{epsfig}
\usepackage{amsmath, slashed}
\usepackage{xcolor}
\usepackage{appendix}
\usepackage{dsfont}

\usepackage[normalem]{ulem}

\usepackage{bm}
\usepackage{amsfonts}
\allowdisplaybreaks
\begin{document}

\title{Twist-3 contribution in the Drell-Yan process with tensor-polarized deuteron}

\author{Si-Yi Qiao}
\affiliation{School of Physics, Zhengzhou University, Zhengzhou, Henan 450001, China}

\author{Qin-Tao Song}
\email[]{songqintao@zzu.edu.cn}
\affiliation{School of Physics, Zhengzhou University, Zhengzhou, Henan 450001, China}
\affiliation{Southern Center for Nuclear-Science Theory (SCNT), Institute of Modern Physics, Chinese Academy of Sciences, Huizhou 516000, China}

\date{\today}

\begin{abstract}
{ The tensor-polarized structures of the deuteron can be probed through the proton-deuteron Drell-Yan process, where the proton is unpolarized and the deuteron is tensor polarized. This measurement will be conducted at Fermilab in the near future. In this reaction, the twist-3 contribution is not negligible compared to the twist-2 contribution due to the limited invariant mass of the dilepton pair.
We calculate the twist-3 contribution for the Drell-Yan cross section with a tensor-polarized deuteron target, preserving the U(1)-gauge invariance of the hadronic tensor.
The cross sections and weighted cross sections are expressed in terms of the tensor-polarized parton distribution functions (PDFs), thus one can extract the PDFs $f_{1\scriptscriptstyle{LL}}$, $f_{\scriptscriptstyle{LT}}$, and  $f^{(1)}_{\scriptscriptstyle{1LT}}$ from the experimental measurements of Drell-Yan process. Our study should be helpful to solve the puzzle in the tensor-polarized structures of the deuteron.}

\end{abstract}

\maketitle

\date{}

\section{Introduction}
\label{introduction}

There are both unpolarized and vector-polarized parton distribution functions (PDFs) in spin-1/2 nucleons, and these PDFs have been intensely studied through the experimental measurements and theoretical work over the past decades. For a  spin-1 hadron, there also exist tensor-polarized PDFs in addition to the unpolarized and vector-polarized ones. The studies of the tensor-polarized PDFs will help us to reveal the inner
  structures of spin-1 hadrons such as deuteron.

The  deuteron is often considered as a weakly bound system of spin-1/2 proton and neutron, in which the proton and neutron are primarily in an S-wave state with only a very tiny D-wave component, and this is known as the standard model of deuteron. There are no tensor-polarized PDFs in nucleons,  and the existence of tensor-polarized structures is related to the orbital angular momentum between proton and neutron in the deuteron. In 2005, the twist-2 tensor-polarized distribution $b_1(x)$ [or $f_{1\scriptscriptstyle{LL}}(x)$]~\cite{Hoodbhoy:1988am, Frankfurt:1983qs} was measured by the HERMES collaboration~\cite{HERMES:2005pon},  however, the experimental measurement of $b_1(x)$ is much larger than the theoretical prediction in Ref.~\cite{Cosyn:2017fbo} where the standard model of deuteron is used. This may suggest that the deuteron cannot be explained by such a simple model. Thus, one can infer that there could be non-nucleonic structures such  as $\Delta \Delta$ and  hidden color components in deuteron~\cite{Miller:2013hla}. Since the uncertainties of the HERMES measurements are large, it is necessary to  measure $b_1(x)$ accurately to solve the puzzle of the tensor-polarized structures in deuteron, and these measurements will be conducted at the Thomas Jefferson National Accelerator Facility (JLab)~\cite{jlab-b1,jlab-b2}, Fermi National Accelerator Laboratory (Fermilab) \cite{fermilab1039,Keller:2020wan,Keller:2022abm,Clement:2023eun}, and Nuclotron-based Ion Collider fAcility (NICA)\cite{Arbuzov:2020cqg}  in the near future. Meanwhile,  the authors of Ref.~\cite{Sargsian:2022rmq} also proposed to use the high-momentum transfer electrodisintegration process to study the non-nucleonic components in the deuteron.

The gluon transversity $\Delta_{\scriptscriptstyle{T}}g(x)$~\cite{Jaffe:1989xy, Nzar:1992ax} is another twist-2 tensor-polarized PDF in a spin-1 hadron. It flips the helicities of the incoming and outgoing gluons, similar to the case of quark transversity.
The gluon transversity of deuteron can be measured by  deep-inelastic scattering (DIS)~\cite{Jaffe:1989xy, Ma:2013yba} and Drell-Yan process~\cite{Kumano:2019igu,Kumano:2020gfk} with a tensor-polarized deuteron target. For a counterpart to PDFs,  one can find the quark collinear fragmentation functions (FFs) up to twist 4  in Ref.~\cite{Ji:1993vw} for a spin-1 hadron.
There are several intriguing theoretical relations for the tensor-polarized PDFs and FFs, including the sum rule of $\int dx b_1(x)=0$~\cite{Close:1990zw}, Wandzura-Wilczek (WW) type relations~\cite{Kumano:2021fem} and Lorentz invariance relations~\cite{Kumano:2021xau,Song:2023ooi}.
In addition to the one-dimensional distributions, there are also studies on the three-dimensional distributions for spin-1 hadrons,  such as  generalized parton distributions (GPDs)~\cite{Cosyn:2018rdm,Berger:2001zb, Shi:2023oll,Zhang:2022zim, Sun:2018ldr, Adhikari:2018umb, Sun:2017gtz,Kumar:2019eck,Taneja:2011sy}, transverse-momentum dependent (TMD) PDFs~\cite{Bacchetta:2000jk,Bacchetta:2002xd,kumano:2021,Ninomiya:2017ggn,Boer:2016xqr} and FFs~\cite{Chen:2016moq,Chen:2023kqw}.
See Ref.~\cite{Kumano:2024fpr} for a recent review on the tensor-polarized structures of spin-1 hadrons.

In the near future, $b_1(x)$ will be measured using the Drell-Yan process with a tensor-polarized deuteron target, where the 120 GeV unpolarized proton beam is provided by the Fermilab Main Injector.
In this reaction, the typical value for the
invariant mass of dilepton pair
is about $Q \sim 4-6$ GeV, which means that the higher-twist contribution can not be neglected.
In the case of the proton-proton Drell-Yan process, the higher-twist effects or single spin asymmetries (SSAs) have been well studied in Refs.~\cite{Qiu:1991pp,Hammon:1996pw,Boer:1997bw, Boer:1999si,Boer:2001tx,Ma:2003ut,Anikin:2010wz,Zhou:2010ui, ReCalegari:2013tnb,Ma:2011nd,Ma:2014uma, Chen:2016dnp, Hu:2021naj,Ma:2012ph} using the collinear factorization.
However, there are only studies on the leading-twist cross section for the proton-deuteron Drell-Yan process at present~\cite{Hino:1998ww,Kumano:2016ude}. In this work we will calculate the twist-3 contribution in the proton-deuteron Drell-Yan process. On one hand, incorporating higher-twist contributions into the cross section will provide a more accurate description of experimental measurements at Fermilab considering the kinematics, and help address the puzzle of tensor-polarized structures in the deuteron. On the other hand, the higher-twist tensor-polarized PDFs such as $f_{\scriptscriptstyle{LT}} (x)$
can also be extracted through the twist-3 analysis of the experimental cross section.

This paper is organized as follows. In Sec.~\ref{PDFs-def}, the intrinsic, kinematical, and dynamical (quark-gluon-quark) correlation functions are
parametrized  in terms of collinear PDFs up to twist 3 for spin-1 hadrons.
We calculate the hadronic tensor of Drell-Yan process with a tensor-polarized deuteron in Sec.~\ref{t3cro}. The cross sections are discussed in Sec.~\ref{t4cro}, and they are expressed  in terms of twist-2 and twist-3 PDFs. A brief summary of this work is given in Sec.~\ref{summary}.

\section{Parton distribution functions in spin-1 hadrons }
\label{PDFs-def}
The polarization vectors $\epsilon$ are used to describe the spin of a pure spin-1 hadron, and they  indicate three spin projections ($\pm1,0$) along a given direction.
If we consider a system of spin-1 hadrons which are not pure states, one need to use the spin density matrix $\rho$ to describe the polarization, and the density matrix is expressed in terms of  the spin vector $S$ and the tensor $T$. The former also exists in the spin-1/2 hadrons known as the vector polarization, however, the tensor polarization $T$ only exist in hadrons  with spin larger than or equal to one.
 The subscripts $L$ and $T$ are commonly used in the literature to denote longitudinal polarization $S_{\scriptscriptstyle{L}}$
and transverse polarization $S_{\scriptscriptstyle{T}}$ for vector-polarized hadrons, respectively.
In this work, we follow the conventions of  Refs.~\cite{Bacchetta:2000jk, Boer:2016xqr} to describe the tensor polarization, where the subscripts $LL$, $LT$, and $TT$
represent different types of tensor polarization.  For tensor-polarized hadrons in the rest frame, the spin density matrix $\rho$  is
parametrized  in terms of the spin parameters  $S_{\scriptscriptstyle{LL}}$, $S_{\scriptscriptstyle{LT}}^{i}$ and $S_{\scriptscriptstyle{TT}}^{ij}$. We use $| m_{(\theta, \phi)} \, \rangle $ to denote the eigenstate with spin projection $m$ along the direction $\vec{v}=( \sin \theta \cos \phi, \sin\theta \sin \phi, \cos \theta )$ for a spin-1 hadron, and the probability of finding this state in a system of spin-1 hadrons can be given by
\begin{align}
 P(m_{(\theta, \phi)})=\text{Tr} \left [ \rho | m_{(\theta, \phi)} \, \rangle    \langle  m_{(\theta, \phi)} \,|   \right].
\label{eqn:probabi}
\end{align}
Then, one can interpret the spin parameters associated with tensor polarization as~\cite{Bacchetta:2000jk}
\begin{align}
&S_{\scriptscriptstyle{LL}}= \frac{P(1_{(0, 0)}) + P(-1_{(0, 0)})}{2}- P(0_{(0, 0)}), \,  S_{\scriptscriptstyle{LT}}^x= P(0_{(-\pi/4, 0)})- P(0_{(\pi/4, 0)}), \,
 S_{\scriptscriptstyle{LT}}^y= P(0_{(-\pi/4, \pi/2)})- P(0_{(\pi/4, \pi/2)})  \nonumber \\
 & S_{\scriptscriptstyle{TT}}^{xx}= P(0_{(\pi/2, \pi/2)})- P(0_{(\pi/2, 0)}),\,
 S_{\scriptscriptstyle{TT}}^{xy}= P(0_{(\pi/2, -\pi/4)})- P(0_{(\pi/2, \pi/4)}) .
\label{eqn:probabi}
\end{align}
If we choose a coordinate frame in which the hadron momentum $P$ is aligned with the $z$ axis, and the covariant form of the matrix $T^{\mu \nu}$ turns out to be
\begin{align}
T^{\mu\nu} & = \frac{1}{2} \left [ \frac{4}{3} S_{\scriptscriptstyle{LL}} \frac{(P^+)^2}{M^2}
                \bar{n}^\mu \bar{n}^\nu
          - \frac{2}{3} S_{\scriptscriptstyle{LL}} (  n^{\{ \mu} \bar n^{\nu \}} -g_{\scriptscriptstyle{T}}^{\mu\nu} )
              + \frac{1}{3} S_{\scriptscriptstyle{LL}} \frac{M^2}{(P^+)^2} n^\mu n^\nu
 + \frac{P^+}{M}  \bar{n}^{\{ \mu}_{\vphantom{T}} S_{\scriptscriptstyle{LT}}^{\nu \}}
- \frac{M}{2 P^+}  n^{\{ \mu}_{\vphantom{T}} S_{\scriptscriptstyle{LT}}^{\nu \}}
+ S_{\scriptscriptstyle{TT}}^{\mu\nu} \right ],
\label{eqn:tensor}
\\[-0.90cm] \nonumber
\end{align}
where $a^{\left\{ \mu \right.}b^{ \left. \nu \right\} } =a^{\mu}b^{\nu}+a^{\nu}b^{\mu}$ denotes symmetrization of the indices,
and $M$ is the hadron mass.
The lightcone  vectors $n$ and $\bar{n}$ are given by
\begin{align}
n^\mu =\frac{1}{\sqrt{2}} (\, 1,\, 0,\, 0,\,  -1 \, ), \ \
\bar n^\mu =\frac{1}{\sqrt{2}} (\, 1,\, 0,\, 0,\,  1 \, ),
\label{eqn:lightcone-n-nbar}
\end{align}
thus, $P$ can be expressed as $P=P^+ \bar{n} +\frac{M^2}{2 P^+ }n$. The lightcone components $a^{\pm}$ of  a Lorentz vector $a^{\mu}$ are defined by
\begin{align}
a^{+}=a\cdot n, \, \,   a^{-}=a\cdot \bar{n}, \, \, a^{\mu}=\left[a^+, a^-, a_{\scriptscriptstyle{T}}^{\mu} \right],
\label{eqn:trans}
\end{align}
and its  transverse component is given by $ a_{\scriptscriptstyle{T}}^{\mu}=g_{\scriptscriptstyle{T}}^{\mu \nu}a_{\nu}$
using the projector onto the subspace orthogonal to $n$ and $\bar{n}$,
\begin{align}
g_{\scriptscriptstyle{T}}^{\mu \nu}=g^{\mu \nu}-n^{\mu}\bar{n}^{\nu}-n^{\nu}\bar{n}^{\mu}.
\label{eqn:trans-sub}
\end{align}
For convenience, the transverse Levi-Civita tensor $\epsilon_{\scriptscriptstyle{T}}^{\alpha \beta } $ is also defined by
\begin{align}
\epsilon_{\scriptscriptstyle{T}}^{\alpha \beta}=\epsilon^{\alpha \beta \mu \nu}  \bar{n}_{\mu}n_{\nu}
\label{eqn:trasn-ten}
\end{align}
with the convention $\epsilon^{0123}=1$.

For tensor-polarized spin-1 hadrons, the collinear correlation function is expressed as
\begin{align}
\Phi_{ij} (x, P, T )
= \int  \frac{d\xi^-}{2\pi} \, e^{ixP^+ \xi^-}
\langle \, P , T \left | \,
\bar\psi _j (0)\mathcal{L}_n(0)  \, \mathcal{L}^{\dag}_n(\xi^-)
\psi _i (\xi^-)  \, \right | \! P, \,  T \,
\rangle,
\label{eqn:correlation-pdf}
\end{align}
where $\mathcal{L}_n(\xi^-)$ is the gauge link to make sure that the gauge invariance is satisfied,
\begin{align}
\mathcal{L}_n(\xi^-)=P \exp[-i g\int^0_{-\infty}dy^-A^+(\xi^-+y^-)].
\label{eqn:gauge}
\end{align}
The correlation function is parametrized  in terms of PDFs~\cite{Frankfurt:1983qs,Hoodbhoy:1988am,kumano:2021, Ma:2013yba, Bacchetta:2000jk},
 \begin{align}
\Phi (x,P,T)
= \frac{1}{2} \bigg\{\Big[\slashed{\bar n} \, f_{1}(x)+ \frac{M}{P^+} \, e(x)\Big] +    \Big[
S_{\scriptscriptstyle{LL}} \, \slashed{\bar n} \, f_{1\scriptscriptstyle{LL}} (x)
+ \frac{M}{P^+} \, S_{\scriptscriptstyle{LL}} \, e_{\scriptscriptstyle{LL}} (x)
+ \frac{M}{P^+} \, \slashed{S}_{\scriptscriptstyle{LT}} \, f_{\scriptscriptstyle{LT}} (x)
\Big ] \bigg\},
\label{eqn:co-pdfs}
\end{align}
where the twist-4 PDFs are neglected since they will be not used in this work, and we also suppress the dependence on the renormalization scale. The first two terms of Eq.~\eqref{eqn:co-pdfs} are unpolarized PDFs, and the rest are related to tensor polarization. For the twist-2 PDF $f_{1\scriptscriptstyle{LL}}$, the notation $b_1=-3/2 f_{1\scriptscriptstyle{LL}}$ was also used in literature,  with the factor of $-3/2$ arising from a different definition of
$S_{\scriptscriptstyle{LL}}$~\cite{Bacchetta:2002xd,Hoodbhoy:1988am}.
Note that there are also vector-polarized PDFs in spin-1 hadrons, similar to the case of spin-1/2 hadrons, but they are omitted here.

In addition, the quark-gluon-quark correlation function is also defined by
\begin{align}
(\Phi_G^\alpha)_{ij} (x_1, x_2)=
\int  \! \frac{d \xi_1^-}{2\pi}   \frac{d \xi_2^-}{2\pi}
 \,    e^{i x_1 P^+ \xi_1^-}  e^{i (x_2-x_1) P^+ \xi_2^-}
\langle \, P, T \left | \,  \bar\psi _j (0) \,
g \, G^{+ \alpha}( \xi_2^- ) \, \psi _i (\xi_1^-)  \,
  \right | P, T \, \rangle,
\label{eqn:3pr}
\end{align}
where the gauge links are suppressed, and the index $\alpha$ is transverse.
At leading twist, it can be expressed in terms of real distribution functions~\cite{Kumano:2021fem, Ma:2013yba},
\begin{align}
\Phi_G^\alpha (x_1, x_2) = \frac{M}{2} \bigg\{  &i \gamma^{\alpha} \slashed{\bar n}E(x_1,x_2)+
 \Big [i S_{\scriptscriptstyle{LT}}^\alpha  \slashed{\bar n}  F_{\scriptscriptstyle{LT}}(x_1, x_2)
- \tilde{S}_{\scriptscriptstyle{LT}}^{\alpha}
\gamma_5 \slashed{\bar n}  G_{\scriptscriptstyle{LT}}(x_1, x_2)
\nonumber \\&
+ i S_{\scriptscriptstyle{LL}} \gamma^{\alpha} \slashed{\bar n}  H_{\scriptscriptstyle{LL}}^\perp (x_1, x_2)
+ i S_{\scriptscriptstyle{TT}}^{\alpha \mu} \gamma_{\mu} \slashed{\bar n}  H_{\scriptscriptstyle{TT}}(x_1, x_2) \Big ] \bigg\}.
\label{eqn:3prede}
\end{align}
where the convention $ \tilde{S}_{\scriptscriptstyle{LT}}^{\alpha}= \epsilon_{\scriptscriptstyle{T}}^{\alpha \mu} S_{\scriptscriptstyle{LT}}\phantom{}_{\mu}$ is used, and  we neglect vector-polarized distributions since they are irrelevant in this work. The arguments $x_1$ and $x_2$ represent the momentum fractions carried by quarks.
In the case of $x_1=x_2$, the momentum fraction carried by gluon vanishes, and the distribution functions are known as the soft gluonic poles. Hermiticity indicates that only $F_{\scriptscriptstyle{LT}}(x_1, x_2)$ is antisymmetric under the exchange of variables $x_1$ and $x_2$, so the soft gluonic pole does not exist in $F_{\scriptscriptstyle{LT}}(x_1, x_2)$, however, the rest functions are symmetric.
The soft gluonic poles have been under intense investigation over the past decades.
For instance, the transverse SSA in proton-proton Drell-Yan process is described using the soft gluonic poles $T_F(x, x)$ (also known as Efremov-Teryaev-Qiu-Sterman matrix element)~\cite{Qiu:1991pp,Qiu:1991wg,Efremov:1984ip} and $T^{(\sigma)}_F(x, x)$ under the twist-3 collinear factorization.
The QCD evolutions for $T_F(x, x)$  and $T^{(\sigma)}_F(x, x)$ are discussed in Refs.~\cite{Vogelsang:2009pj,Kang:2008ey,Zhou:2008mz,Braun:2009mi,Schafer:2012ra, Kang:2012em, Ma:2012xn,Ma:2012ye}.
In addition,  $T_F(x, x)$  and $T^{(\sigma)}_F(x, x)$ are related to the transverse moments of time-reversal-odd Silvers function and Boer-Mulders function, respectively~\cite{Zhou:2010ui,Boer:2003cm, Ma:2003ut}.
We also parametrize~\cite{Bacchetta:2000jk}
\begin{align}
&\int \frac{d\xi^{\,-}}{2 \pi} \, e^{ix P^+ \xi^{\,-}} \, \langle \, P, T \, | \, \bar{\psi}_j(0) \, i \partial^{\alpha} \,\psi_i(\xi^{\,-}) \, | \, P, T \, \rangle \nonumber \\
=&\frac{M}{2} \left \{ i h_{1}^{\perp  (1)}(x) \gamma_{\scriptscriptstyle{T}}^{\alpha} \, \slashed{\bar n}+
\left[  f_{1\scriptscriptstyle{LT}}^{(1)} (x) \, S_{\scriptscriptstyle{LT}}^{\alpha} \,
    \slashed{\bar n}
    + g_{1\scriptscriptstyle{LT}}^{(1)} (x) \,
\tilde{S}_{\scriptscriptstyle{LT}}^{\alpha} \, \gamma_5 \, \slashed{\bar n}
- h_{1\scriptscriptstyle{LL}}^{\perp  (1)} (x)
  S_{\scriptscriptstyle{LL}} \sigma^{\alpha \mu} \bar n_{\mu}
+ h_{1\scriptscriptstyle{TT}}^{\prime   (1)} (x)
 \, S_{\scriptscriptstyle{TT}}^{\alpha \beta} \,
        \sigma_{\beta \mu} \, \bar n^{\,\mu}  \right] \right \},
\label{eqn:kt-co1}
\end{align}
where the vector-polarized distributions and the gauge links are neglected.

We have now defined three types of twist-3 distributions, which can be classified following the conventions of Ref.~\cite{Kanazawa:2015ajw}. Specifically,
the twist-3 distributions in Eqs.~\eqref{eqn:co-pdfs}, \eqref{eqn:3prede}, and \eqref{eqn:kt-co1} are referred to as intrinsic, dynamical, and kinematical distributions, respectively. Note that these distributions are not entirely independent, as
the relations among intrinsic,
kinematical, and dynamical distributions can be derived using the QCD equation of motion (for example, see Eq.~\eqref{eqn:ten-eom}).

\section{Twist-3 contribution to the hadronic tensor in  Drell-Yan process}
\label{t3cro}

\begin{figure}[ht]
\includegraphics[angle=0,scale=0.4]{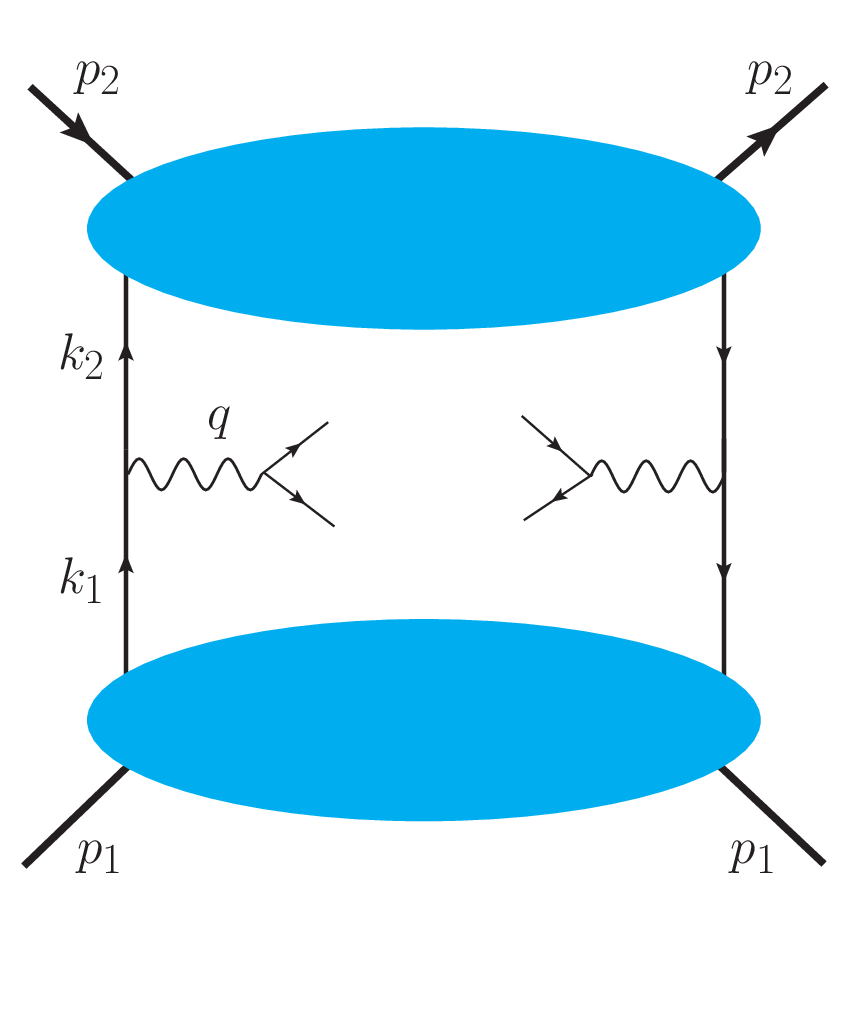}
\caption{
Feynman diagram of proton-deuteron Drell-Yan process, where only the quark-quark correlation functions are included.}
	\label{fig01}
\end{figure}

There are numerous studies on the twist-3 contribution in the proton-proton Drell-Yan process under the collinear factorization. In this work, we will adopt the method that was developed in Refs.~\cite{Ma:2014uma, Chen:2016dnp, Hu:2021naj} to investigate the proton-deuteron Drell-Yan process using Feynman gauge,  and its measurements will be conducted at Fermilab with a tensor-polarized deuteron.
The hadronic tensor of the Drell-Yan process is calculated in the center-of-mass frame for deuteron and proton, where the momentum of deuteron is chosen to be the $z$ axis.
In Fig.~\ref{fig01}, $k_1$ denotes the parton momentum that comes from deuteron, and $k_2$ is the parton momentum from the proton. The power counting indicates that $k_1^+$  and $k_1^-$ are of order $Q$, which is the invariant mass of the dilepton pair. Thus, we can have
\begin{align}
k_1^{\mu} \sim  Q \left[1, \lambda^2, \lambda, \lambda\right],   k_2^{\mu} \sim Q \left[\lambda^2, 1, \lambda, \lambda \right]
\label{eqn:twt}
\end{align}
with $\lambda \sim \Lambda_{\text{QCD}}/Q$. In the collinear factorization, the quark-gluon-quark distributions are also included in the hadronic tensor to obtain the twist-3 contribution, and
we expand the hadronic tensor in terms of the power of $\lambda$.
Although the hadronic tensor is calculated in the center-of-mass frame, it is expressed in terms of the hadron momenta which are Lorentz vectors.
We contract the hadronic tensor with the lepton tensor to obtain the cross section, and the angular distribution of the lepton pair will be shown in the Collins-Soper frame~\cite{Collins:1977iv}.

In the following, we will show how to  calculate the twist-3 cross section for the proton-deuteron Drell-Yan process,
 \begin{align}
d(p_1)+p(p_2)\to \mu^-(l_1)\mu^+(l_2)+X,
\label{eqn:dy}
\end{align}
where proton is unpolarized and  the deuteron is polarized. In general, the cross section contains three parts, and they are the unpolarized, vector-polarized, and tensor polarized ones. The first two are similar as the cross section of the proton-proton Drell-Yan process, which have already been studied in the literature, and the tensor-polarized part is defined by
\begin{align}
d\hat{\sigma}=\frac{d\sigma(T)-d\sigma(-T)}{2}.
\label{eqn:cro-t}
\end{align}
The cross section  is expressed in terms of the   hadronic tensor $W^{\mu\nu}$ and the lepton tensor $L_{\mu\nu}$,
\begin{align}
\frac{d\hat{\sigma}}{dQ^2d\Omega}=\frac{\alpha^2e_q^2}{4SQ^4}\int d^4q \delta(q^2-Q^2) W^{\mu\nu}L_{\mu\nu},
\label{eqn:cro}
\end{align}
where $\Omega$ is the solid angle of the lepton pair. $S$ and $q$ is defined by
\begin{align}
S=(p_1+p_2)^2, \, \, q=l_1+l_2, \, \, Q^2=q^2.
\end{align}

\begin{figure}[ht]
	\includegraphics[angle=0,scale=0.4]{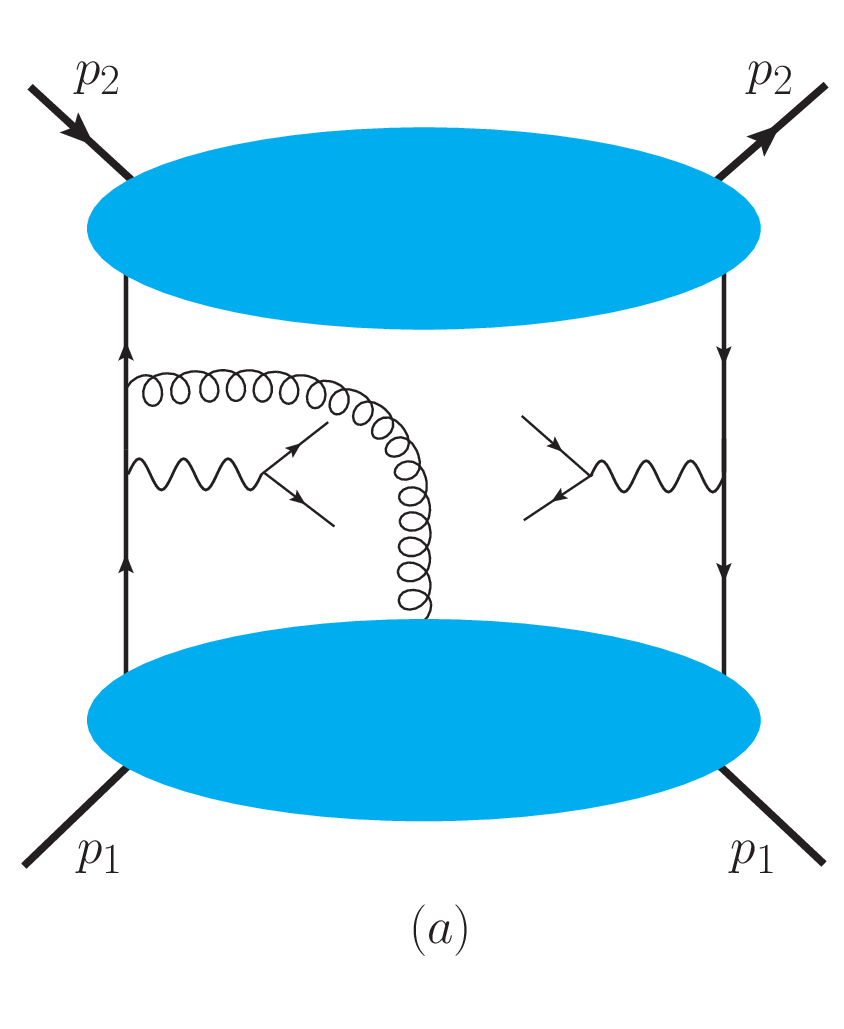}\hspace{0.75cm}
 \includegraphics[angle=0,scale=0.4]{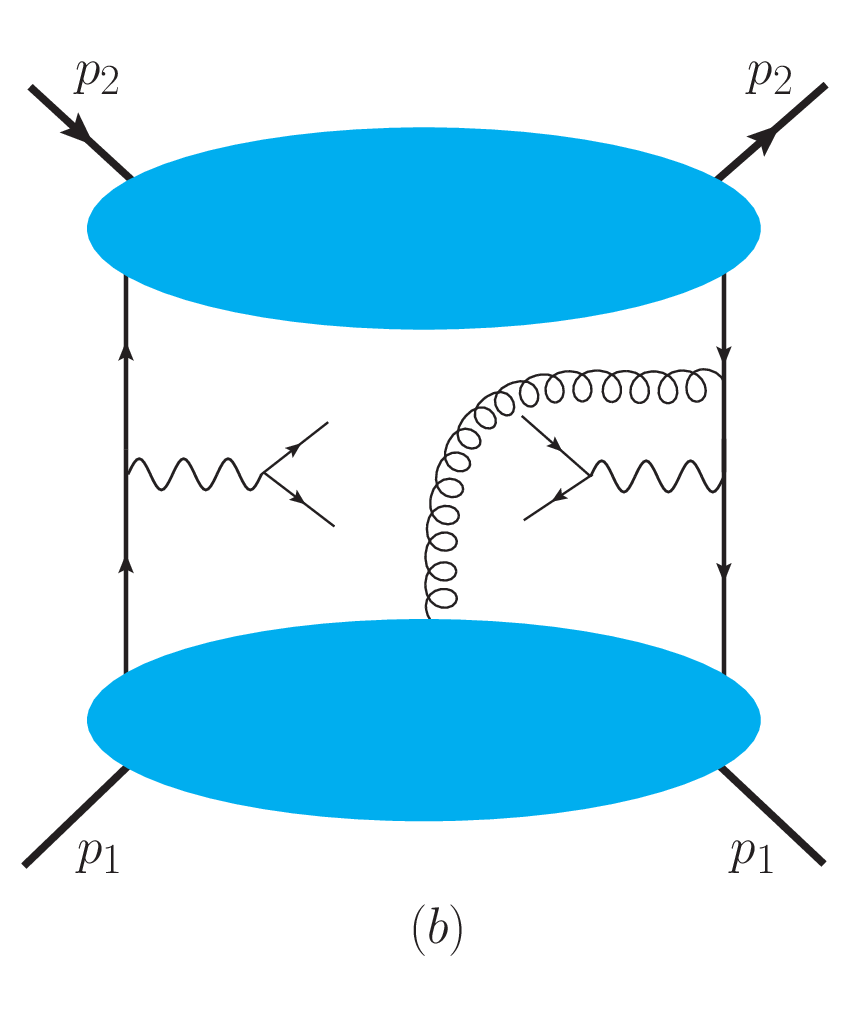}
	\caption{Feynman diagrams of proton-deuteron Drell-Yan process with one gluon emitted from the lower bubble.}
	\label{fig02}
\end{figure}

To obtain the twist-3 contribution in the Drell-Yan process, we  need to calculate the   hadronic tensor at twist 3.
In Fig.~\ref{fig01}, the   hadronic tensor is expressed in terms of quark-quark correlation functions, and it is given by
\begin{align}
 W_1^{\mu\nu}=\frac{1}{2N_c} \int d^3 k_1 dk_2^- f_1^{\bar{q}}(y) \left[ \delta^4(q-k_1-k_2)(\gamma^{\mu} \gamma^+ \gamma^{\nu})_{ji} \right] \int \frac{d^3\bar{\xi}}{(2\pi)^3} e^{i\xi \cdot k_1 }
 \langle \, p_1  \left | \,
\bar\psi _j (0) \, \psi _i (\xi)  \, \right | \! p_1\rangle,
\label{eqn:hten1}
\end{align}
where $N_c=3$ is the color factor, and we use the following notations,
\begin{align}
 \xi^{\mu}=\left[0, \xi^-, \xi_{\scriptscriptstyle{T}}^{\mu}\right], \, \, d^3k_1=dk_1^+d^2k_{1\scriptscriptstyle{T}}, \, \,d^3\bar{\xi}=d\xi^-d^2\xi_{\scriptscriptstyle{T}}.
\label{eqn:mom}
\end{align}
In Eq.~\eqref{eqn:hten1}, we need to take the transverse moment of  $k_1$ into account,
\begin{align}
k_1^{\mu}=\left[k_1^+, 0, k_{1\scriptscriptstyle{T}}^{\mu}\right], \, \, k_2^{\mu}=\left[0, k_2^-, 0\right],\,\, k_1^+=xp_1^+, \,\, k_2^-=yp_2^-,
\label{eqn:mom2}
\end{align}
and one can also consider the case of  $k_{2\scriptscriptstyle{T}} \neq0$ for the twist-3 contribution; however, this contribution vanishes since the proton is unpolarized. The $\gamma$ matrices in Eq.~\eqref{eqn:hten1} can be expanded into
\begin{align}
\gamma^{\mu} \gamma^+ \gamma^{\nu}=\left(- g_{\scriptscriptstyle{T}}^{\mu \nu}\gamma^++n^{\{\mu}_{\vphantom{\scriptscriptstyle{T}}} \gamma^{\nu\}}_{\scriptscriptstyle{T}} \right)-
i\left(\epsilon_{\scriptscriptstyle{T}}^{\mu \nu} \gamma_5 \gamma^+ + n^{[\mu}_{\vphantom{\scriptscriptstyle{T}}}\epsilon_{\scriptscriptstyle{T}}^{\nu]\rho} \gamma_5 \gamma_{\scriptscriptstyle{T}}\phantom{}_{ \rho}   \right)+2n^{\mu}n^{\nu} \gamma^-,
\label{eqn:3gamma}
\end{align}
where the convention $a^{\left[ \mu \right.}b^{ \left. \nu \right] } =a^{\mu}b^{\nu}-a^{\nu}b^{\mu}$ is used. The last term is beyond twist 3, and it can be neglected in the calculation.
Then, Eq.~\eqref{eqn:3gamma} can be divided into the vector part and axial-vector part. The latter gives rise to the antisymmetric part of the hadronic tensor, which does not contribute to the cross section after being contracted with the lepton tensor. The twist-2 contribution in the  hadronic tensor is obtained using the $\gamma^+$ term of Eq.~\eqref{eqn:3gamma}, and it is  expressed as~\cite{Hino:1998ww}
\begin{align}
 W_{1}^{\mu\nu}\Big |_{\text{twist 2}}=- \frac{1}{N_c}\delta^2(q_{\scriptscriptstyle{T}})f^{\bar{q}}_1(y) f^q_{1\scriptscriptstyle{LL}}(x)S_{\scriptscriptstyle{LL}}g_{\scriptscriptstyle{T}}^{\mu\nu},
\label{eqn:ten-vt2}
\end{align}
where  $q$ is the  quark flavor.
Note that the gauge links in the distributions $f^q_{1\scriptscriptstyle{LL}}(x)$  and $f^{\bar{q}}_1(y)$  are obtained by including the $A^+$ and $A^-$ gluon emissions from the lower and higher bubbles, respectively.
As discussed in Sec.~\ref{PDFs-def}, the twist-3 corrections can be categorized into intrinsic, kinematical, and dynamical types.
In Fig.~\ref{fig01}, the twist-3 contribution comes from the $\gamma_{\scriptscriptstyle{T}}$ term of Eq.~\eqref{eqn:3gamma} and the expansion of  transverse moment in the $\delta$ function of Eq.~\eqref{eqn:hten1}. This contribution is given by intrinsic and kinematical distributions,
\begin{align}
 W_{1}^{\mu\nu}\Big |_{\text{twist 3}}= \frac{M}{N_c}    f^{\bar{q}}_1(y)   \left[ \delta^2(q_{\scriptscriptstyle{T}})   \frac{p^{\{\mu}_{2}S_{\scriptscriptstyle{LT}}^{\nu\}} }{p_1\cdot p_2}
f^q_{\scriptscriptstyle{LT}}(x)
+g_{\scriptscriptstyle{T}}^{\mu\nu} \frac{\partial \delta^2(q_{\scriptscriptstyle{T}}) }{\partial q_{\scriptscriptstyle{T}}^{\rho} } S_{\scriptscriptstyle{LT}}^{\rho} f^{(1)q}_{\scriptscriptstyle{1LT}}(x)
  \right],
\label{eqn:ten-vt3}
\end{align}
and we can see  that this expression does not satisfy the $U(1)$-gauge invariance, which is different from the twist-2 contribution.
This is because that we have not considered the contribution that is expressed in term of dynamical quark-gluon-quark distributions, and they are shown in Fig.~\ref{fig02} where one gluon is emitted from lower bubble. We can also include the gluon emission from the higher bubble, and it does not contribute to the  hadronic tensor, which is similar as the expansion of the parton's transverse moment in Eq.~\eqref{eqn:mom2}.

For the contribution in Fig.~\ref{fig02}a, the   hadronic tensor is given by
\begin{align}
 W_{2a}^{\mu\nu}=\frac{1}{2N_c} \int d^3 k_1  d^3 k  dk_2^- f_1^{\bar{q}}(y)  \delta^4(q-k_1-k_2) \text{Tr} \left[\gamma_{\rho} \frac{\slashed{k}+\slashed{k}_2}{(k+k_2)^2+i \epsilon} \gamma^{\nu} \Psi_{\scriptscriptstyle{A}}^{\rho}(k_1-k,k_1) \gamma^{\mu} \gamma^+ \right],
\label{eqn:hten2}
\end{align}
where $\Psi_{\scriptscriptstyle{A}}^{\rho}(k_1,k_1-k)$ is the matrix element of the quark-gluon-quark operator,
\begin{align}
\Psi^{\alpha}_{\scriptscriptstyle{A}ij}(k_1-k,k_1)= \int \frac{d^3\bar{\xi} d^3\bar{\xi}_1 }{(2\pi)^6} e^{i\xi_1 \cdot k_1+ i(\xi-\xi_1 )\cdot k }
 \langle \, p_1 \left | \,
\bar\psi _j (0) \, g A^{\alpha}(\xi) \, \psi _i (\xi_1)  \, \right | \! p_1\rangle \big |_{\xi^+=\xi_1^+=0} .
\label{eqn:dis3}
\end{align}
 Similarly, the contribution from Fig.~\ref{fig02}b can be written as
\begin{align}
 W_{2b}^{\mu\nu}=\frac{1}{2N_c} \int d^3 k_1  d^3 k  dk_2^- f_1^{\bar{q}}(y)  \delta^4(q-k_1-k_2) \text{Tr}
  \left[ \gamma^{\mu} \frac{\slashed{k}+\slashed{k}_2}{(k+k_2)^2-i \epsilon} \gamma_{\rho} \gamma^+\gamma^{\nu}   \Psi_{\scriptscriptstyle{A}}^{\rho}(k_1,k_1-k)     \right].
\label{eqn:hten3}
\end{align}
We neglect the contributions  that are beyond the order of $\lambda$ in Eqs.~\eqref{eqn:hten2} and \eqref{eqn:hten3}, and the  hadronic  tensor contributed by Fig.~\ref{fig02} is given by the dynamical quark-gluon-quark distributions,
\begin{align}
 W^{\mu\nu}_{2}=- \frac{M}{N_c}f^{\bar{q}}_1(y)\delta^2(q_T)       \frac{p^{\{\mu}_{1}S_{\scriptscriptstyle{LT}}^{\nu\}} }{y p_1\cdot p_2} \,          \mathcal{P} \int dy_1 \frac{F^q_{\scriptscriptstyle{LT}}(x,y_1)+ G^q_{\scriptscriptstyle{LT}}(x,y_1) }{x-y_1}
\label{eqn:ten-tab},
\end{align}
where $\mathcal{P}$ indicates the principal integral. The dynamical quark-gluon-quark distributions can be replaced by the intrinsic and kinematical distributions thanks to
the QCD equation of motion for quarks~\cite{Kumano:2021xau},
\begin{align}
x f^q_{\scriptscriptstyle{LT}}(x)-f^{(1)q}_{\scriptscriptstyle{1LT}}(x)
- \mathcal{P} \int dy_1 \frac{F^q_{\scriptscriptstyle{LT}}(x,y_1)+ G^q_{\scriptscriptstyle{LT}}(x,y_1) }{x-y_1}=0.
\label{eqn:ten-eom}
\end{align}
We add the higher-twist contributions of Eqs.~\eqref{eqn:ten-vt3} and  \eqref{eqn:ten-tab} to Eq.~\eqref{eqn:ten-vt2}, and the complete twist-3 correction to the   hadronic tensor can be expressed as
\begin{align}
 W^{\mu\nu}= \frac{1}{N_c} f^{\bar{q}}_1(y) \bigg\{ - \delta^2(q_{\scriptscriptstyle{T}}) S_{\scriptscriptstyle{LL}}g_{\scriptscriptstyle{T}}^{\mu\nu} f^q_{1\scriptscriptstyle{LL}}(x) +&
 \frac{M}{p_1\cdot p_2}  \delta^2(q_{\scriptscriptstyle{T}})   \left[   p^{\{\mu}_{2}S_{\scriptscriptstyle{LT}}^{\nu\}}f^q_{\scriptscriptstyle{LT}}(x)-
 p^{\{\mu}_{1}S_{\scriptscriptstyle{LT}}^{\nu\}}\frac{1}{y} \left( x f^q_{\scriptscriptstyle{LT}}(x)-f^{(1)q}_{\scriptscriptstyle{1LT}}(x)\right)
  \right] \nonumber \\
+&M g_{\scriptscriptstyle{T}}^{\mu\nu} \frac{\partial \delta^2(q_{\scriptscriptstyle{T}}) }{\partial q_{\scriptscriptstyle{T}}^{\rho} } S_{\scriptscriptstyle{LT}}^{\rho} f^{(1)q}_{\scriptscriptstyle{1LT}}(x)
 \bigg\}
\label{eqn:ten-sum},
\end{align}
where the first term is leading-twist contribution that is associated with the $S_{\scriptscriptstyle{LL}}$-type tensor polarization. The rest ones are twist-3 effects, and they appear in the case of the $S_{\scriptscriptstyle{LT}}$-type tensor polarization.
In the proton-proton Drell-Yan process, the transverse SSA is given by the soft gluonic poles, which can be regarded as time-reversal-odd distributions, as discussed in Sec.~\ref{PDFs-def}.
However,  there is no twist-3 contribution originating  from the soft gluonic poles in Eq.~\eqref{eqn:ten-sum}. Moreover, the twist-3 distributions $f_{\scriptscriptstyle{LT}}$ and $f^{(1)}_{\scriptscriptstyle{1LT}}$ are time-reversal-even functions.

The   hadronic tensor of Eq.~\eqref{eqn:ten-sum} should be considered as a distribution of $q_{\scriptscriptstyle{T}}$ due to $\delta^2(q_{\scriptscriptstyle{T}})$, thus, the $U(1)$-gauge invariance can be checked only after the integration over $q_{\scriptscriptstyle{T}}$~\cite{Ma:2014uma},
\begin{align}
 \int d^2q_{\scriptscriptstyle{T}}W^{\mu\nu}q_{\mu} \mathcal{F}(q_{\scriptscriptstyle{T}})=0
\label{eqn:ten-gauge},
\end{align}
where $\mathcal{F}(q_T)$ is a test function of $q_{\scriptscriptstyle{T}}$. The expression of Eq.~\eqref{eqn:ten-sum} satisfies the $U(1)$-gauge invariance in Eq.~\eqref{eqn:ten-gauge}.

\section{Cross section of   Drell-Yan process}
\label{t4cro}

Now we will check the cross section in the Collins-Soper frame where $\mu^+ \mu^-$ is at rest, and the momentum of $\mu^-$ is expressed as
\begin{align}
l_1^{\mu}=\frac{Q}{2}(1, \sin\theta \cos \phi,\sin\theta \sin \phi, \cos \theta ).
\label{eqn:mum}
\end{align}
The hadron polarization is often defined in the laboratory frame or the center-of-mass frame of deuteron and proton, and the transverse vector $S_{\scriptscriptstyle{LT}}^{\mu}$  is given by
\begin{align}
S_{\scriptscriptstyle{LT}}^{\mu}=|S_{\scriptscriptstyle{LT}}|(0, \cos \phi_s, \sin \phi_s, 0).
\label{eqn:st}
\end{align}
The vector $S_{\scriptscriptstyle{LT}}^{\mu}$ should be different from the laboratory frame to  the Collins-Soper frame due to the  Lorentz boost. However, this difference is at order of $|q_{\scriptscriptstyle{T}}|^2$, and one can safely neglect it.

\begin{figure}[htb]
\includegraphics[angle=0,scale=0.55]{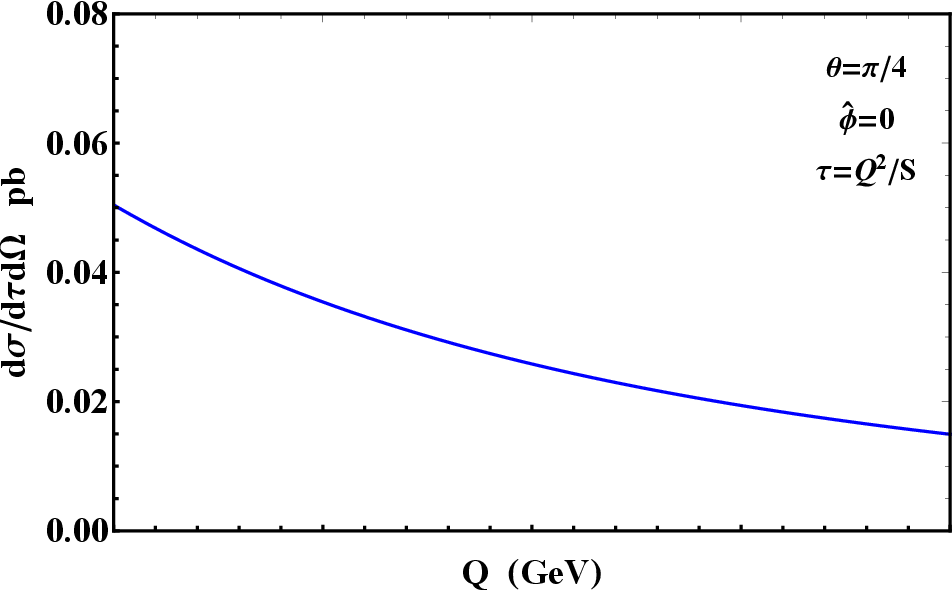}
\caption{The twist-3  cross section for the proton-deuteron Drell-Yan process at Fermilab.}
	\label{fig03}
\end{figure}

The differential cross section is obtained after one contracts the   hadronic tensor $W^{\mu \nu}$ with the lepton tensor $L^{\mu \nu}$, and it is expressed as
\begin{align}
\frac{d\hat{\sigma}}{dQ^2d\Omega}=\sum_q \frac{\alpha^2e_q^2}{4N_cQ^2}\int dx dy  \delta(xyS-Q^2)   \bigg\{& S_{\scriptscriptstyle{LL}}
\left[ f^q_{\scriptscriptstyle{1LL}}(x) f^{\bar{q}}_1(y) +(q \leftrightarrow \bar{q})   \right](1+\cos^2\theta) \nonumber \\
+&|S_{\scriptscriptstyle{LT}}| \frac{M}{Q}  \left[ (2xf^q_{\scriptscriptstyle{LT}}(x)- f^{(1)q}_{\scriptscriptstyle{1LT}}(x)) f^{\bar{q}}_1(y)+(q \leftrightarrow \bar{q})  \right] \sin(2\theta)\cos\hat{\phi}
\bigg\},
\label{eqn:crofi}
\end{align}
where $\hat{\phi}=\phi-\phi_s$. In the case of the $S_{\scriptscriptstyle{LL}}$-type tensor polarization, the leading-twist PDF $f_{1\scriptscriptstyle{LL}}$ can be extracted from the cross section.
The twist-3 PDFs $f_{\scriptscriptstyle{LT}}$ and $f^{(1)}_{\scriptscriptstyle{1LT}}$ appear in the twist-3 part of cross section with  the $S_{\scriptscriptstyle{LT}}$-type polarized deuteron, and they are suppressed by $M/Q$.
One can roughly estimate that the twist-3 correction is approximately one-third of the leading-twist cross section, considering the kinematics of the Drell-Yan process at Fermilab. However, this estimate is valid under the assumption that $|S_{\scriptscriptstyle{LL}}| \sim |S_{\scriptscriptstyle{LT}}|$.
Since  $|S_{\scriptscriptstyle{LL}}|$ can be significantly smaller than $|S_{\scriptscriptstyle{LT}}|$ for certain polarization states of the deuteron target, the twist-3 contribution will dominate the tensor-polarized cross section in these cases.
At present, $f_{\scriptscriptstyle{LT}}$ and $f^{(1)}_{\scriptscriptstyle{1LT}}$ remain unknown in Eq.~\eqref{eqn:crofi}, making it challenging to calculate the twist-3 correction precisely.
Previous theoretical and experimental studies indicate that tensor-polarized PDFs are much smaller than unpolarized and vector-polarized ones in the deuteron.
However, one can assume that the twist-2 and twist-3 tensor-polarized PDFs are of a similar magnitude, namely, $|f_{\scriptscriptstyle{LT}}| \sim |f^{(1)}_{\scriptscriptstyle{1LT}}| \sim  |f_{\scriptscriptstyle{1LL}}|$.
In Fig.~\ref{fig03}, we provide a rough numerical estimate of the higher-twist correction based on the assumption $2xf^q_{\scriptscriptstyle{LT}}(x)- f^{(1)q}_{\scriptscriptstyle{1LT}}(x)= f^{q}_{\scriptscriptstyle{1LL}}(x)$, using the extracted PDF $f^{q}_{\scriptscriptstyle{1LL}}(x)$  from the Hermes measurement~\cite{Kumano:2010vz}.
 The range of $Q$ is chosen as
 $ 4$ GeV$ \leq Q \leq 6$ GeV, which is typical for the proton-deuteron Drell-Yan process at Fermilab.
The polar and azimuthal angles are fixed at $\theta = \pi/4$ and $\hat{\phi} = 0$, respectively, with the spin parameter set to $|S_{\scriptscriptstyle{LT}}| = 1$.
Our results indicate that the higher-twist correction is of the order of 0.1 pb for the Drell-Yan process at Fermilab.   It is important to note that this is merely an order-of-magnitude estimate due to the unknown twist-3 PDFs, and the actual measurement in the near future could reveal significant deviations.

Integrating Eq.~\eqref{eqn:crofi} over the solid angle of $\mu^+ \mu^-$, we can obtain
\begin{align}
\frac{d\hat{\sigma}}{dQ^2}=S_{\scriptscriptstyle{LL}}\sum_q \frac{4 \pi \alpha^2e_q^2}{3N_cQ^2}\int dx dy  \delta(xyS-Q^2)
\left[ f^q_{\scriptscriptstyle{1LL}}(x) f^{\bar{q}}_1(y) +(q \leftrightarrow \bar{q})   \right],
\label{eqn:crofi1}
\end{align}
where only  $f_{1\scriptscriptstyle{LL}}$ is involved. To separate $f_{\scriptscriptstyle{LT}}$ from $f^{(1)}_{\scriptscriptstyle{1LT}}$ in Eq.~\eqref{eqn:crofi}, one can also define the weighted differential cross section,
\begin{align}
\frac{d\hat{\sigma}\langle \mathcal{F}_1 \rangle }{dQ^2d\Omega}=\frac{\alpha^2e_q^2}{4SQ^4}\int d^4q \delta(q^2-Q^2) W^{\mu\nu}L_{\mu\nu} \mathcal{F}_1(q_{\scriptscriptstyle{T}}),
\label{eqn:crowe}
\end{align}
where $\mathcal{F}_1(q_{\scriptscriptstyle{T}})=q_{\scriptscriptstyle{T}} \cdot S_{\scriptscriptstyle{LT}}/Q$ is defined. Then, the differential cross section is expressed in terms of $f^{(1)}_{\scriptscriptstyle{1LT}}$,
\begin{align}
\frac{d\hat{\sigma}\langle \mathcal{F}_1 \rangle }{dQ^2}=-|S_{\scriptscriptstyle{LT}}|^2 \frac{4\pi \alpha^2e_q^2M}{3N_cQ^3}\int dx dy  \delta(xyS-Q^2) \left[ f^{(1)q}_{\scriptscriptstyle{1LT}}(x) f^{\bar{q}}_1(y)+  (q \leftrightarrow \bar{q}) \right].
\label{eqn:crowe2}
\end{align}

Since the unpolarized PDF $f_1$ of proton can be considered as a well-known quantity, one can extract the tensor-polarized  PDFs $f_{1\scriptscriptstyle{LL}}$, $f_{\scriptscriptstyle{LT}}$, and  $f^{(1)}_{\scriptscriptstyle{1LT}}$ from the cross sections of Eqs.~\eqref{eqn:crofi}, \eqref{eqn:crofi1}, and \eqref{eqn:crowe2}.
Although the dynamical quark-gluon-quark distributions do not explicitly appear in the cross sections, they can be probed through the combination of the two twist-3 PDFs,  $x f_{\scriptscriptstyle{LT}}-f^{(1)}_{\scriptscriptstyle{1LT}}$, as shown in Eq.~\eqref{eqn:ten-eom}.
For the leading-twist PDF $f_{1\scriptscriptstyle{LL}}$, it has been extracted from the HERMES measurements with slightly large  uncertainties~\cite{Kumano:2010vz}. Furthermore, we can also use the extracted PDFs to
test the WW-type relation~\cite{Kumano:2021fem},
\begin{align}
f_{\scriptscriptstyle{LT}}(x)=\frac{3}{2} \int_x^1 \frac{dy}{y}f_{1\scriptscriptstyle{LL}}(y).
\label{eqn:ww}
\end{align}
The expression of Eq.~\eqref{eqn:ww} is analogous to the WW relation between the vector-polarized PDFs $g_1$ and $g_2$ (or $g_T$) in proton~\cite{Wandzura:1977qf}.
Recent study has shown that the violation of the WW relation can be as large as 15\%-40\% of the magnitude of  $g_2$~\cite{Accardi:2009au}, and this violation is attributed to dynamical quark-gluon-quark distributions.

\section{Summary}
\label{summary}

Recently, the tensor-polarized PDFs of the spin-1 hadrons
have drawn a lot of attention in the hadronic physics. Studying these PDFs provides valuable insight into the internal structures of spin-1 hadrons. For instance, the structure function $b_1(x)$ [or $f_{1\scriptscriptstyle{LL}}(x)$] would vanish in the deuteron if its internal proton and neutron were purely in the S-wave state.  However, HERMES data indicate a significant discrepancy between the experimental measurements and the theoretical predictions based on the standard model of the deuteron,  thus, there could be non-nucleonic structures  in the deuteron~\cite{Miller:2013hla}.
Because of the large uncertainties in previous measurements, physicists have proposed accurate measurements of $b_1(x)$
at various experimental facilities to solve the puzzle of the tensor-polarized structures in the deuteron.

At Fermilab, $f_{1\scriptscriptstyle{LL}}$  will be probed through the proton-deuteron Drell-Yan process with a tensor-polarized deuteron. Given that the invariant mass of dilepton pair is only a few GeV, the twist-3 correction should play an important role in the measured cross section.
In this work,  we calculate the twist-3 contribution to the Drell-Yan process,  providing a more accurate description of the future measurements at Fermilab.
The tensor-polarized PDF $f_{1\scriptscriptstyle{LL}}$ appears only in the twist-2 part of the cross section.
If we include the twist-3 correction into the cross section, the tensor-polarized
PDFs $f_{\scriptscriptstyle{LT}}$ and  $f^{(1)}_{\scriptscriptstyle{1LT}}$ can be extracted from the experimental measurements in addition to
$f_{1\scriptscriptstyle{LL}}$.
Thus, one can test the WW-type relation between $f_{\scriptscriptstyle{LT}}$ and  $f_{1\scriptscriptstyle{LL}}$.
In addition, we provide a rough numerical estimate of the twist-3 correction for the Drell-Yan process at Fermilab.
Our study also applies to the pion-deuteron Drell-Yan process, and this measurement may be possible at COMPASS in the future.

\section*{Acknowledgments}
We acknowledge useful discussions with Shunzo Kumano, Bernard Pire, Guang-Peng Zhang, and Jian Zhou.
Qin-Tao Song was supported by the National Natural Science Foundation
of China under Grant Number 12005191.




\begin{thebibliography}{00}

















































\bibitem{Hoodbhoy:1988am}
P.~Hoodbhoy, R.~L.~Jaffe and A.~Manohar,
Nucl. Phys. B \textbf{312} (1989), 571-588.

\bibitem{Frankfurt:1983qs}
L.~L.~Frankfurt and M.~I.~Strikman,
Nucl. Phys. A \textbf{405} (1983), 557-580.



\bibitem{HERMES:2005pon}
A.~Airapetian \textit{et al.} [HERMES],
Phys. Rev. Lett. \textbf{95} (2005), 242001.




\bibitem{Cosyn:2017fbo}
W.~Cosyn, Y.~B.~Dong, S.~Kumano and M.~Sargsian,
Phys. Rev. D \textbf{95} (2017) no.7, 074036.






\bibitem{Miller:2013hla}
G.~A.~Miller,
Phys. Rev. C \textbf{89} (2014) no.4, 045203.




\bibitem{jlab-b1} J.-P. Chen {\it et al.},
The Deuteron Tensor Structure Function $b_1$, Proposal to Jefferson Lab PAC-38, PR12-11-110 (2011).

\bibitem{jlab-b2}
M. Jones {\it et al.}, Search for Exotic Gluonic States in the Nucleus, A Letter of Intent to Jefferson Lab PAC 44, LOI12-16-006 (2016), arXiv:1803.11206 [nucl-ex].


\bibitem{fermilab1039}
M. Brooks,\textit{et al.}, SeaQuest with a Transversely Polarized Target (E1039).





\bibitem{Keller:2020wan}
D.~Keller, D.~Crabb and D.~Day,
Nucl. Instrum. Meth. A \textbf{981} (2020), 164504.



\bibitem{Keller:2022abm}
D.~Keller [SpinQuest],
[arXiv:2205.01249 [nucl-ex]].

\bibitem{Clement:2023eun}
J.~Clement and D.~Keller,
Nucl. Instrum. Meth. A \textbf{1050} (2023), 168177.









\bibitem{Arbuzov:2020cqg}
A.~Arbuzov, A.~Bacchetta, M.~Butenschoen, F.~G.~Celiberto, U.~D'Alesio, M.~Deka, I.~Denisenko, M.~G.~Echevarria, A.~Efremov and N.~Y.~Ivanov, \textit{et al.}
Prog. Part. Nucl. Phys. \textbf{119} (2021), 103858.


\bibitem{Sargsian:2022rmq}
M.~M.~Sargsian and F.~Vera,
Phys. Rev. Lett. \textbf{130} (2023) no.11, 112502.















\bibitem{Jaffe:1989xy}
R.~L.~Jaffe and A.~Manohar,
Phys. Lett. B \textbf{223} (1989), 218-224.





\bibitem{Nzar:1992ax}
M.~Nzar and P.~Hoodbhoy,
Phys. Rev. D \textbf{45} (1992), 2264-2268.









\bibitem{Ma:2013yba}
J.~P.~Ma, C.~Wang and G.~P.~Zhang,
[arXiv:1306.6693 [hep-ph]].



\bibitem{Kumano:2019igu}
S.~Kumano and Q.~T.~Song,
Phys. Rev. D \textbf{101} (2020) no.5, 054011.


\bibitem{Kumano:2020gfk}
S.~Kumano and Q.~T.~Song,
Phys. Rev. D \textbf{101} (2020) no.9, 094013.


\bibitem{Close:1990zw}
F.~E.~Close and S.~Kumano,
Phys. Rev. D \textbf{42} (1990), 2377-2379.


















\bibitem{Kumano:2021fem}
S.~Kumano and Q.~T.~Song,
JHEP \textbf{09} (2021), 141.


\bibitem{Kumano:2021xau}
S.~Kumano and Q.~T.~Song,
Phys. Lett. B \textbf{826} (2022), 136908.


\bibitem{Song:2023ooi}
Q.~T.~Song,
Phys. Rev. D \textbf{108} (2023) no.9, 094041.





\bibitem{Ji:1993vw}
X.~D.~Ji,
Phys. Rev. D \textbf{49} (1994), 114-124.



\bibitem{Berger:2001zb}
E.~R.~Berger, F.~Cano, M.~Diehl and B.~Pire,
Phys. Rev. Lett. \textbf{87} (2001), 142302.


\bibitem{Cosyn:2018rdm}
W.~Cosyn and B.~Pire,
Phys. Rev. D \textbf{98} (2018) no.7, 074020.




\bibitem{Sun:2018ldr}
B.~D.~Sun and Y.~B.~Dong,
Phys. Rev. D \textbf{99} (2019) no.1, 016023.



\bibitem{Adhikari:2018umb}
L.~Adhikari, Y.~Li, M.~Li and J.~P.~Vary,
Phys. Rev. C \textbf{99} (2019) no.3, 035208.



\bibitem{Sun:2017gtz}
B.~D.~Sun and Y.~B.~Dong,
Phys. Rev. D \textbf{96} (2017) no.3, 036019.


\bibitem{Kumar:2019eck}
N.~Kumar,
Phys. Rev. D \textbf{99} (2019) no.1, 014039.


\bibitem{Taneja:2011sy}
S.~K.~Taneja, K.~Kathuria, S.~Liuti and G.~R.~Goldstein,
Phys. Rev. D \textbf{86} (2012), 036008.







\bibitem{Shi:2023oll}
C.~Shi, J.~Li, P.~L.~Yin and W.~Jia,
Phys. Rev. D \textbf{107} (2023) no.7, 074009.



\bibitem{Zhang:2022zim}
J.~L.~Zhang, G.~Z.~Kang and J.~L.~Ping,
Phys. Rev. D \textbf{105} (2022) no.9, 094015.



















\bibitem{Bacchetta:2000jk}
A.~Bacchetta and P.~J.~Mulders,
Phys. Rev. D \textbf{62} (2000), 114004.

\bibitem{Bacchetta:2002xd}
A.~Bacchetta,
[arXiv:hep-ph/0212025 [hep-ph]].




\bibitem{Boer:2016xqr}
D.~Boer, S.~Cotogno, T.~van Daal, P.~J.~Mulders, A.~Signori and Y.~J.~Zhou,
JHEP \textbf{10} (2016), 013.


\bibitem{Ninomiya:2017ggn}
Y.~Ninomiya, W.~Bentz and I.~C.~Clo\"et,
Phys. Rev. C \textbf{96} (2017) no.4, 045206.




\bibitem{kumano:2021}
S. Kumano and Qin-Tao Song,
Phys. Rev. D \textbf{103} (2021), 014025.




































\bibitem{Chen:2016moq}
K.~B.~Chen, W.~H.~Yang, S.~Y.~Wei and Z.~T.~Liang,
Phys. Rev. D \textbf{94} (2016) no.3, 034003.

\bibitem{Chen:2023kqw}
K.~B.~Chen, T.~Liu, Y.~K.~Song and S.~Y.~Wei,
Particles \textbf{6} (2023) no.2, 515-545.



\bibitem{Kumano:2024fpr}
S.~Kumano,
Eur. Phys. J. A \textbf{60} (2024) no.10, 205.









\bibitem{Qiu:1991pp}
J.~W.~Qiu and G.~F.~Sterman,
Phys. Rev. Lett. \textbf{67} (1991), 2264-2267.



\bibitem{Hammon:1996pw}
N.~Hammon, O.~Teryaev and A.~Schafer,
Phys. Lett. B \textbf{390} (1997), 409-412.



\bibitem{Boer:1997bw}
D.~Boer, P.~J.~Mulders and O.~V.~Teryaev,
Phys. Rev. D \textbf{57} (1998), 3057-3064.


\bibitem{Boer:1999si}
D.~Boer and P.~J.~Mulders,
Nucl. Phys. B \textbf{569} (2000), 505-526.



\bibitem{Boer:2001tx}
D.~Boer and J.~W.~Qiu,
Phys. Rev. D \textbf{65} (2002), 034008.


\bibitem{Ma:2003ut}
J.~P.~Ma and Q.~Wang,
Eur. Phys. J. C \textbf{37} (2004), 293-298.



\bibitem{Anikin:2010wz}
I.~V.~Anikin and O.~V.~Teryaev,
Phys. Lett. B \textbf{690} (2010), 519-525.


\bibitem{Zhou:2010ui}
J.~Zhou and A.~Metz,
Phys. Rev. D \textbf{86} (2012), 014001.

\bibitem{ReCalegari:2013tnb}
G.~Re Calegari and P.~G.~Ratcliffe,
Eur. Phys. J. C \textbf{74} (2014), 2769.


\bibitem{Ma:2011nd}
J.~P.~Ma and H.~Z.~Sang,
JHEP \textbf{04} (2011), 062.



\bibitem{Ma:2012ph}
J.~P.~Ma and G.~P.~Zhang,
JHEP \textbf{11} (2012), 156.




\bibitem{Ma:2014uma}
J.~P.~Ma and G.~P.~Zhang,
JHEP \textbf{02} (2015), 163
[arXiv:1409.2938 [hep-ph]].


\bibitem{Chen:2016dnp}
A.~P.~Chen, J.~P.~Ma and G.~P.~Zhang,
Phys. Rev. D \textbf{95} (2017) no.7, 074005.




















\bibitem{Hu:2021naj}
M.~C.~Hu, J.~P.~Ma, Z.~Y.~Pang and G.~P.~Zhang,
Phys. Rev. D \textbf{105} (2022) no.1, 014009.










\bibitem{Hino:1998ww}
S.~Hino and S.~Kumano,
Phys. Rev. D \textbf{59} (1999), 094026.



\bibitem{Kumano:2016ude}
S.~Kumano and Q.~T.~Song,
Phys. Rev. D \textbf{94} (2016) no.5, 054022.


















\bibitem{Qiu:1991wg}
J.~W.~Qiu and G.~F.~Sterman,
Nucl. Phys. B \textbf{378} (1992), 52-78.






\bibitem{Efremov:1984ip}
A.~V.~Efremov and O.~V.~Teryaev,
Phys. Lett. B \textbf{150} (1985), 383.



\bibitem{Vogelsang:2009pj}
W.~Vogelsang and F.~Yuan,
Phys. Rev. D \textbf{79} (2009), 094010.



\bibitem{Kang:2008ey}
Z.~B.~Kang and J.~W.~Qiu,
Phys. Rev. D \textbf{79} (2009), 016003.

\bibitem{Zhou:2008mz}
J.~Zhou, F.~Yuan and Z.~T.~Liang,
Phys. Rev. D \textbf{79} (2009), 114022.

\bibitem{Braun:2009mi}
V.~M.~Braun, A.~N.~Manashov and B.~Pirnay,
Phys. Rev. D \textbf{80} (2009), 114002
[erratum: Phys. Rev. D \textbf{86} (2012), 119902].

\bibitem{Schafer:2012ra}
A.~Schafer and J.~Zhou,
Phys. Rev. D \textbf{85} (2012), 117501.



\bibitem{Kang:2012em}
Z.~B.~Kang and J.~W.~Qiu,
Phys. Lett. B \textbf{713} (2012), 273-276


\bibitem{Ma:2012xn}
J.~P.~Ma and Q.~Wang,
Phys. Lett. B \textbf{715} (2012), 157-163.



\bibitem{Ma:2012ye}
J.~P.~Ma, Q.~Wang and G.~P.~Zhang,
Phys. Lett. B \textbf{718} (2013), 1358-1363.





\bibitem{Boer:2003cm}
D.~Boer, P.~J.~Mulders and F.~Pijlman,
Nucl. Phys. B \textbf{667} (2003), 201-241



\bibitem{Kanazawa:2015ajw}
K.~Kanazawa, Y.~Koike, A.~Metz, D.~Pitonyak and M.~Schlegel,
Phys. Rev. D \textbf{93} (2016) no.5, 054024.







\bibitem{Collins:1977iv}
J.~C.~Collins and D.~E.~Soper,
Phys. Rev. D \textbf{16} (1977), 2219.



\bibitem{Kumano:2010vz}
S.~Kumano,
Phys. Rev. D \textbf{82} (2010), 017501.






\bibitem{Wandzura:1977qf}
S.~Wandzura and F.~Wilczek,
Phys. Lett. B \textbf{72} (1977), 195-198.





\bibitem{Accardi:2009au}
A.~Accardi, A.~Bacchetta, W.~Melnitchouk and M.~Schlegel,
JHEP \textbf{11} (2009), 093.


















































\end{thebibliography}
\end{document}